\documentclass{article} 
\usepackage{iclr2025_conference,times}

\usepackage{amsmath,amsfonts,bm}

\def\eqref#1{equation~\ref{#1}}

\def\1{\bm{1}}

\DeclareMathAlphabet{\mathsfit}{\encodingdefault}{\sfdefault}{m}{sl}
\SetMathAlphabet{\mathsfit}{bold}{\encodingdefault}{\sfdefault}{bx}{n}

 \iclrfinalcopy
\usepackage{hyperref}
\usepackage{url}
\usepackage{cite}

\title{Opportunities and Challenges of Frontier Data Governance With Synthetic Data}

\author{Madhavendra Thakur \\
Independent Researcher \\
\texttt{mt3890@columbia.edu} \\
\And
Jason Hausenloy \\
University of California, Berkeley \\
\texttt{hausenloy@berkeley.edu} \\
}
\begin{document}

\maketitle

\begin{abstract}
Synthetic data, or data generated by machine learning models, is increasingly emerging as a solution to the data access problem. However, its use introduces significant governance and accountability challenges, and potentially debases existing governance paradigms, such as compute and data governance. In this paper, we identify 3 key governance and accountability challenges that synthetic data poses - it can enable the increased emergence of malicious actors, spontaneous biases and value drift. We thus craft 3 technical mechanisms to address these specific challenges, finding applications for synthetic data towards adversarial training, bias mitigation and value reinforcement. These could not only counteract the risks of synthetic data, but serve as critical levers for governance of the frontier in the future.
\end{abstract}

\section{Introduction}
The rapid advancement of AI has led to an impending data bottleneck, where frontier models require exponentially increasing volumes of high-quality data, with some suggesting that the size of training corpora may exceed the sum of all human-generated data by 2030 \citep{villalobos2024rundatalimitsllm}. Empirical model capability scaling laws dictate that this scarcity bounds the overall capabilities of the frontier, regardless of strides in algorithmic complexity or computational power \citep{ruan2024observationalscalinglawspredictability}, thus positioning it as a key issue for model developers to address. In the long term, data-efficient architectures may emerge in response to this problem and come to define the frontier. However, in the short term, it appears that synthetic data, or data generated by machine learning models as opposed to by humans, is increasingly defining the frontier. Indeed, already, a large portion of the training data of leading models is synthetic \citep{openai2024gpt4technicalreport}.

This shift towards synthetic data is one of many evolutions in the way models are trained that may jeopardize the efficacy of current approaches to governing and ensuring trustworthiness on the frontier. Compute governance, as proposed by Sastry et. al., for example, regulates computational power as a proxy for model capabilities \citep{sastry2024computingpowergovernanceartificial}, but the link between the two grows more tenuous in the face of compute-efficient architectures and distilled models \citep{deepseekai2025deepseekr1incentivizingreasoningcapability}. Similarly, data governance, as proposed by Hausenloy et. al., relies partially on governing the flow of data through "AI Data Supply Chain" \citep{hausenloy2024datagovernancefrontierai}, which they note becomes degenerate with regards to synthetic data, where the data generators, processors and trainers are the same people.

However, we propose that the advent of synthetic data, instead of being a limit to governance efforts, offers unique opportunities as a regulatory lever in addition to its challenges. Indeed, these may serve to be critical vectors for model control in the future, as existing approaches grow less effective, as detailed above.

Specifically, we identify 3 key challenges that synthetic data may pose to governance and accountability initiatives, and craft technical mechanisms to not only counter them, but establish synthetic data as a robust lever for governance of the frontier.

\section{Related Work}

This paper is in the style of similar proposals for governance paradigms, such as the aforementioned "compute governance" and "data governance" \citep{sastry2024computingpowergovernanceartificial, hausenloy2024datagovernancefrontierai}. Specifically, it builds upon this work by addressing a technical and temporal hole in the previous paradigms - how they will apply to synthetic data specifically, and the future more generally.

Our work lies within the subfield of Technical AI Governance \citep{reuel2024openproblemstechnicalai}, as its primary application is translational: proposing technical mechanisms for policy applications, while simultaneously providing a roadmap for technical and policy work from a governance perspective.

\section*{Challenges}

We outline 3 key challenges synthetic data poses to governance and accountability frameworks

\subsection{Synthetic data can be used to generate misaligned data at scale.}

The same ability to produce vast, tailored examples that makes synthetic data attractive to model trainers makes synthetic data attractive to malicious actors. Instead of using traditional, transparent data pipelines~\citep{steinhardt2017certified}, adversaries can mass-produce skewed data to deliberately misinform models~\citep{biggio2012poisoning,jagielski2018manipulating}. Without proper safeguards, such “data poisoning” or “value hijacking” can lead to harmful ideologies or unreliable predictions in critical sectors like healthcare, finance, or public policy~\citep{bender2021stochastic,carlini2019evaluating}.

\subsection{Synthetic data can detach models from real-world contexts.}

Rich synthetic environments, while useful for scalable training, risk insulating models from the dynamic value signals present in authentic human interactions. Without continual exposure to genuine linguistic subtleties, cultural norms, and ethical considerations, models may develop value systems that diverge from societal expectations~\citep{shrivastava2017learning,torralba2011unbiased}. This insulation is further exacerbated by feedback loops where models are retrained on their own synthetic outputs, potentially entrenching misaligned values over time~\citep{richter2020can,ganin2016domain}.

\subsection{Synthetic data could lead to spontaneous biases in black-box systems.}

When large models are repeatedly retrained on their own synthetic outputs, the inherent opacity of deep learning architectures can allow small biases to accumulate unpredictably~\citep{mehrabi2021survey}. Over time, these biases may distort model outputs and compromise fairness, yielding results that conflict with societal expectations~\citep{caliskan2017semantics,blodgett2020language,bender2021stochastic}.

\section{Opportunities and Mechanisms}

We propose 3 mechanisms to counter the challenges outline above.

\subsection{Synthetic Data for Adversarial Training}

Synthetic data for adversarial training offers a scalable approach to enhance the robustness and safety of large-scale models by systematically generating malicious or deceptive scenarios. This counteracts the challenge of synthetic data for misaligned data generation. These scenarios can be used to identify and correct weaknesses in the model that might be exploited in real-world attacks. By synthetically creating adversarial examples at scale, researchers and practitioners can refine model behavior post-training, ultimately contributing to more secure frontier AI systems.

\textbf{Implementation:} To incorporate synthetic adversarial data in a training pipeline, one first delineates the scope of potential attacks (e.g., specific perturbations, semantic manipulations, or deceptive prompts). A generative model such as a Variational Autoencoder or a diffusion-based generator can then be trained on a seed corpus of real-world examples, introducing adversarial constraints during data generation. This yields a large corpus of synthetic adversarial samples. These samples are integrated into the fine-tuning phase of the model, where iterative testing and updating ensure that previously uncovered weaknesses are addressed. Over multiple training rounds, the model learns to withstand a diverse set of adversarial inputs.

\textbf{Existing work:} Past research on adversarial training has largely relied on perturbations of real data~\citep{goodfellow2014explaining,madry2018towards}, but recent work has shown promise in generating synthetic adversarial inputs using Generative Adversarial Networks (GANs)~\citep{creswell2018generative} or large-scale language models~\citep{brown2020language}. Other studies have leveraged simulation frameworks and domain-specific generative models~\citep{kingma2013auto,dhariwal2021diffusion} to produce highly varied adversarial examples that mimic real-world conditions. These approaches indicate that synthetic data can be a powerful tool in building adversarially robust systems, freeing the model developer from reliance on exhaustively labeled, human-crafted attacks.

\textbf{Challenges and mitigation:} While synthetic adversarial data broadens the space of potential attacks, it may also introduce novel biases if the generation process is insufficiently diverse or guided by incomplete threat models. Maintaining alignment between synthetic data distributions and real-world attack vectors can be difficult, requiring continuous monitoring and updating of generative pipelines. Additionally, the iterative feedback loop—whereby models trained on synthetic adversarial data might in turn generate subsequent synthetic data—demands careful oversight to prevent the accumulation of unrealistic or unrepresentative scenarios. Despite these challenges, synthetic adversarial data remains a valuable strategy for improving model robustness and proactively defending against the evolving landscape of security threats.

\subsection{Synthetic Data for Statistical Distribution Balancing}

\textbf{Motivation:} Real-world training datasets often exhibit numerical imbalances in representation. For example, financial transaction data might predominantly originate from urban centers rather than rural areas, potentially affecting model performance across different geographic regions~\citep{Kumar_2022}. This introduces functional inconsistencies in system performance and can compromise operational reliability, as frontier AI systems develop higher accuracy for regions represented in the training data

\textbf{Implementation:} Synthetic data provides a methodical approach to address these representation gaps by algorithmically generating additional examples where data is sparse. This addresses the challenge of distributional patterns emerging in synthetic data. One technical approach involves utilizing generative models trained on available high-quality samples, then supplementing with carefully engineered synthetic instances~\citep{chawla2002smote}. In financial risk assessment, generative adversarial networks have been utilized to produce synthetic transaction records that represent complex, multi-dimensional attributes~\citep{choi2017generating}. Industry experts should be consulted during the generation process to ensure adherence to established standards and regulatory compliance.

\textbf{Existing work:} Technical literature documents various applications of generative methodologies to address dataset completeness. Research has demonstrated that generating synthetic geological data using advanced generative adversarial networks can improve prediction accuracy across different terrain types~\citep{karras2019style}, and comparable data augmentation techniques have been applied to industrial sensor readings~\citep{wei2019eda}. Additionally, frameworks such as RepresentGAN have been developed specifically to generate data that addresses distributional imbalances in training sets~\citep{xu2019fairgan}.

\textbf{Challenges and mitigation:} While synthetic data offers a technical solution for addressing dataset imbalances, it may unintentionally introduce statistical anomalies or inadequately represent real-world distributions. Overreliance on artificially constructed examples may result in models with diminished performance under complex, real-world conditions. Rigorous testing against actual data remains essential. Furthermore, comprehensive documentation of synthetic data generation—detailing methodological assumptions, constraints, and potential error sources—enables stakeholders to properly evaluate the technical validity and reliability of models trained using such approaches~\citep{mehrabi2021survey}.

\subsection{Synthetic Data for Value Reinforcement}

\textbf{Motivation:} Large-scale AI models are increasingly vulnerable to data poisoning and value hijacking, wherein adversarial actors inject harmful ideologies or manipulative content into open-source training corpora~\citep{biggio2012poisoning,steinhardt2017certified}. Such attacks can distort a model’s values, nudging its decisions or outputs toward harmful agendas. By contrast, synthetic data generation provides an opportunity to purposefully curate the values embedded in a training set. This counteracts the challenge of synthetic data leading to detached environments. Rather than indiscriminately scraping the web---where harmful, misleading, or biased content may dominate~\citep{bender2021stochastic}---lab-curated synthetic corpora can emphasize collaborative, ethical, and socially constructive values.

\textbf{Implementation:} To implement value reinforcement via synthetic data, developers can design generative models or specialized data augmentation pipelines that focus on producing content aligned with a set of predefined principles. For instance, a language model might be guided to generate texts that uphold specific ethical frameworks or emphasize fairness and respect across different cultural perspectives~\citep{ziegler2019fine}. This process can include the following steps:
\begin{enumerate}
    \item \textit{Define Value Targets:} Collaborate with ethicists, domain experts, and stakeholders to outline desirable attributes and behaviors, translating them into clear guidelines for synthetic data generation~\citep{amodei2016concrete}.
    \item \textit{Curated Seed Data:} Compile a smaller, high-quality corpus exemplifying the targeted values. This set serves as the seed for training or fine-tuning a generative model.
    \item \textit{Generative Pipeline:} Employ large language models, diffusion-based methods, or other generative frameworks to produce synthetic samples that faithfully reflect the curated seed’s values. Mechanisms such as reinforcement learning or policy gradients can ensure alignment with these standards~\citep{christiano2017deep}.
    \item \textit{Validation and Iteration:} Validate generated content against established guidelines. Discard or correct any synthetic instances that deviate from the desired value set. Iteratively retrain or fine-tune the model as needed~\citep{gehman2020realtotoxicityprompts}.
\end{enumerate}

\textbf{Incentives for AI Labs:} Beyond ethical considerations, AI developers have pragmatic reasons to invest in value-aligned synthetic data. Models trained on carefully curated content often demonstrate higher-quality outputs, more robust performance, and fewer public-relations liabilities. By proactively filtering out harmful or adversarial material, labs can mitigate reputational risks, reduce moderation overhead, and foster user trust. As a result, curation becomes more than a moral imperative—it is also a strategic advantage.

\textbf{Challenges and mitigation:} Achieving broad consensus on which values to promote can be contentious, particularly when cultural, political, or organizational perspectives diverge. Additionally, overly restrictive curation may limit the model’s exposure to diverse viewpoints, potentially compromising its adaptability or realism. Regular review by multidisciplinary teams can help calibrate the balance between value alignment and open-world robustness. Finally, just as data poisoning can subvert open datasets, sophisticated attackers may attempt to introduce subtle biases into curated pipelines, necessitating continual monitoring, audits, and transparency in the curation process.

\section*{Conclusion}
\paragraph{}Synthetic data offers a powerful yet double-edged solution for frontier AI. It can overcome data scarcity and enhance model robustness, but without proper oversight, it risks fostering misaligned values and entrenched biases. The future of synthetic data in AI governance depends on innovative oversight mechanisms and transparent, collaborative frameworks that ensure its benefits are realized while maintaining technical reliability and operational consistency.

\newpage

\bibliography{iclr2025_conference}

\begin{thebibliography}{35}
\providecommand{\natexlab}[1]{#1}
\providecommand{\url}[1]{\texttt{#1}}
\expandafter\ifx\csname urlstyle\endcsname\relax
  \providecommand{\doi}[1]{doi: #1}\else
  \providecommand{\doi}{doi: \begingroup \urlstyle{rm}\Url}\fi

\bibitem[Amodei et~al.(2016)Amodei, Olah, Steinhardt, Christiano, Schulman, and Mane]{amodei2016concrete}
Dario Amodei, Chris Olah, Jacob Steinhardt, Paul Christiano, John Schulman, and Danny Mane.
\newblock Concrete problems in ai safety.
\newblock Technical report, OpenAI, 2016.

\bibitem[Bender et~al.(2021)Bender, Gebru, McMillan-Major, and Shmitchell]{bender2021stochastic}
Emily~M Bender, Timnit Gebru, Angelina McMillan-Major, and Shmargaret Shmitchell.
\newblock On the dangers of stochastic parrots: Can language models be too big?
\newblock In \emph{Proceedings of the 2021 ACM Conference on Fairness, Accountability, and Transparency}, pp.\  610--623, 2021.

\bibitem[Biggio et~al.(2012)Biggio, Nelson, and Laskov]{biggio2012poisoning}
Battista Biggio, Blaine Nelson, and Pavel Laskov.
\newblock Poisoning attacks against support vector machines.
\newblock In \emph{Proceedings of the 29th International Conference on Machine Learning (ICML-12)}, pp.\  1467--1474, 2012.

\bibitem[Blodgett et~al.(2020)Blodgett, O'Connor, Van~Durme, and Green]{blodgett2020language}
Su~Lin Blodgett, Brendan O'Connor, Benjamin Van~Durme, and Lydia Green.
\newblock Language (technology) is power: A critical survey of “bias” in nlp.
\newblock In \emph{Proceedings of the 2020 Conference on Empirical Methods in Natural Language Processing (EMNLP)}, pp.\  529--544, 2020.

\bibitem[Brown et~al.(2020)Brown, Mann, Ryder, et~al.]{brown2020language}
Tom Brown, Benjamin Mann, Nick Ryder, et~al.
\newblock Language models are few-shot learners.
\newblock In \emph{Advances in Neural Information Processing Systems (NeurIPS)}, 2020.

\bibitem[Caliskan et~al.(2017)Caliskan, Bryson, and Narayanan]{caliskan2017semantics}
Aylin Caliskan, Joanna~J Bryson, and Arvind Narayanan.
\newblock Semantics derived from language corpora contain human-like biases.
\newblock \emph{Science}, 356\penalty0 (6334):\penalty0 183--186, 2017.

\bibitem[Carlini et~al.(2019)]{carlini2019evaluating}
Nicholas Carlini et~al.
\newblock Evaluating and testing adversarial robustness in machine learning.
\newblock In \emph{Proceedings of the IEEE Symposium on Security and Privacy}, 2019.

\bibitem[Chawla et~al.(2002)Chawla, Bowyer, Hall, and Kegelmeyer]{chawla2002smote}
Nitesh~V Chawla, Kevin~W Bowyer, Lawrence~O Hall, and W~Philip Kegelmeyer.
\newblock Smote: Synthetic minority over-sampling technique.
\newblock \emph{Journal of Artificial Intelligence Research}, 16:\penalty0 321--357, 2002.

\bibitem[Choi et~al.(2017)Choi, Biswal, Malin, Duke, Stewart, and Sun]{choi2017generating}
Edward Choi, S~Biswal, Benjamin Malin, John Duke, Walter~F Stewart, and Jimeng Sun.
\newblock Generating multi-label discrete electronic health records using generative adversarial networks.
\newblock In \emph{Machine Learning for Healthcare Conference}, pp.\  286--305, 2017.

\bibitem[Christiano et~al.(2017)Christiano, Leike, Brown, Martic, Legg, and Amodei]{christiano2017deep}
Paul~F Christiano, Jan Leike, Tom~B Brown, Miljan Martic, Shane Legg, and Dario Amodei.
\newblock Deep reinforcement learning from human preferences.
\newblock In \emph{Advances in Neural Information Processing Systems}, volume~30, pp.\  4299--4307, 2017.

\bibitem[Creswell et~al.(2018)Creswell, White, Dumoulin, Arulkumaran, Sengupta, and Bharath]{creswell2018generative}
Antonia Creswell, Tom White, Vincent Dumoulin, Kai Arulkumaran, Bidisha Sengupta, and Anil~A. Bharath.
\newblock Generative adversarial networks: An overview.
\newblock \emph{IEEE Signal Processing Magazine}, 2018.

\bibitem[DeepSeek-AI et~al.(2025)DeepSeek-AI, Guo, Yang, Zhang, Song, Zhang, Xu, Zhu, Ma, Wang, Bi, Zhang, Yu, Wu, Wu, Gou, Shao, Li, Gao, Liu, Xue, Wang, Wu, Feng, Lu, Zhao, Deng, Zhang, Ruan, Dai, Chen, Ji, Li, Lin, Dai, Luo, Hao, Chen, Li, Zhang, Bao, Xu, Wang, Ding, Xin, Gao, Qu, Li, Guo, Li, Wang, Chen, Yuan, Qiu, Li, Cai, Ni, Liang, Chen, Dong, Hu, Gao, Guan, Huang, Yu, Wang, Zhang, Zhao, Wang, Zhang, Xu, Xia, Zhang, Zhang, Tang, Li, Wang, Li, Tian, Huang, Zhang, Wang, Chen, Du, Ge, Zhang, Pan, Wang, Chen, Jin, Chen, Lu, Zhou, Chen, Ye, Wang, Yu, Zhou, Pan, Li, Zhou, Wu, Ye, Yun, Pei, Sun, Wang, Zeng, Zhao, Liu, Liang, Gao, Yu, Zhang, Xiao, An, Liu, Wang, Chen, Nie, Cheng, Liu, Xie, Liu, Yang, Li, Su, Lin, Li, Jin, Shen, Chen, Sun, Wang, Song, Zhou, Wang, Shan, Li, Wang, Wei, Zhang, Xu, Li, Zhao, Sun, Wang, Yu, Zhang, Shi, Xiong, He, Piao, Wang, Tan, Ma, Liu, Guo, Ou, Wang, Gong, Zou, He, Xiong, Luo, You, Liu, Zhou, Zhu, Xu, Huang, Li, Zheng, Zhu, Ma, Tang, Zha, Yan, Ren, Ren, Sha, Fu, Xu, Xie, Zhang,
  Hao, Ma, Yan, Wu, Gu, Zhu, Liu, Li, Xie, Song, Pan, Huang, Xu, Zhang, and Zhang]{deepseekai2025deepseekr1incentivizingreasoningcapability}
DeepSeek-AI, Daya Guo, Dejian Yang, Haowei Zhang, Junxiao Song, Ruoyu Zhang, Runxin Xu, Qihao Zhu, Shirong Ma, Peiyi Wang, Xiao Bi, Xiaokang Zhang, Xingkai Yu, Yu~Wu, Z.~F. Wu, Zhibin Gou, Zhihong Shao, Zhuoshu Li, Ziyi Gao, Aixin Liu, Bing Xue, Bingxuan Wang, Bochao Wu, Bei Feng, Chengda Lu, Chenggang Zhao, Chengqi Deng, Chenyu Zhang, Chong Ruan, Damai Dai, Deli Chen, Dongjie Ji, Erhang Li, Fangyun Lin, Fucong Dai, Fuli Luo, Guangbo Hao, Guanting Chen, Guowei Li, H.~Zhang, Han Bao, Hanwei Xu, Haocheng Wang, Honghui Ding, Huajian Xin, Huazuo Gao, Hui Qu, Hui Li, Jianzhong Guo, Jiashi Li, Jiawei Wang, Jingchang Chen, Jingyang Yuan, Junjie Qiu, Junlong Li, J.~L. Cai, Jiaqi Ni, Jian Liang, Jin Chen, Kai Dong, Kai Hu, Kaige Gao, Kang Guan, Kexin Huang, Kuai Yu, Lean Wang, Lecong Zhang, Liang Zhao, Litong Wang, Liyue Zhang, Lei Xu, Leyi Xia, Mingchuan Zhang, Minghua Zhang, Minghui Tang, Meng Li, Miaojun Wang, Mingming Li, Ning Tian, Panpan Huang, Peng Zhang, Qiancheng Wang, Qinyu Chen, Qiushi Du, Ruiqi Ge, Ruisong
  Zhang, Ruizhe Pan, Runji Wang, R.~J. Chen, R.~L. Jin, Ruyi Chen, Shanghao Lu, Shangyan Zhou, Shanhuang Chen, Shengfeng Ye, Shiyu Wang, Shuiping Yu, Shunfeng Zhou, Shuting Pan, S.~S. Li, Shuang Zhou, Shaoqing Wu, Shengfeng Ye, Tao Yun, Tian Pei, Tianyu Sun, T.~Wang, Wangding Zeng, Wanjia Zhao, Wen Liu, Wenfeng Liang, Wenjun Gao, Wenqin Yu, Wentao Zhang, W.~L. Xiao, Wei An, Xiaodong Liu, Xiaohan Wang, Xiaokang Chen, Xiaotao Nie, Xin Cheng, Xin Liu, Xin Xie, Xingchao Liu, Xinyu Yang, Xinyuan Li, Xuecheng Su, Xuheng Lin, X.~Q. Li, Xiangyue Jin, Xiaojin Shen, Xiaosha Chen, Xiaowen Sun, Xiaoxiang Wang, Xinnan Song, Xinyi Zhou, Xianzu Wang, Xinxia Shan, Y.~K. Li, Y.~Q. Wang, Y.~X. Wei, Yang Zhang, Yanhong Xu, Yao Li, Yao Zhao, Yaofeng Sun, Yaohui Wang, Yi~Yu, Yichao Zhang, Yifan Shi, Yiliang Xiong, Ying He, Yishi Piao, Yisong Wang, Yixuan Tan, Yiyang Ma, Yiyuan Liu, Yongqiang Guo, Yuan Ou, Yuduan Wang, Yue Gong, Yuheng Zou, Yujia He, Yunfan Xiong, Yuxiang Luo, Yuxiang You, Yuxuan Liu, Yuyang Zhou, Y.~X. Zhu,
  Yanhong Xu, Yanping Huang, Yaohui Li, Yi~Zheng, Yuchen Zhu, Yunxian Ma, Ying Tang, Yukun Zha, Yuting Yan, Z.~Z. Ren, Zehui Ren, Zhangli Sha, Zhe Fu, Zhean Xu, Zhenda Xie, Zhengyan Zhang, Zhewen Hao, Zhicheng Ma, Zhigang Yan, Zhiyu Wu, Zihui Gu, Zijia Zhu, Zijun Liu, Zilin Li, Ziwei Xie, Ziyang Song, Zizheng Pan, Zhen Huang, Zhipeng Xu, Zhongyu Zhang, and Zhen Zhang.
\newblock Deepseek-r1: Incentivizing reasoning capability in llms via reinforcement learning, 2025.
\newblock URL \url{https://arxiv.org/abs/2501.12948}.

\bibitem[Dhariwal \& Nichol(2021)Dhariwal and Nichol]{dhariwal2021diffusion}
Prafulla Dhariwal and Alexander~Quinn Nichol.
\newblock Diffusion models beat {GANs} on image synthesis.
\newblock In \emph{Advances in Neural Information Processing Systems (NeurIPS)}, 2021.

\bibitem[Ganin et~al.(2016)Ganin, Ustinova, Ajakan, Germain, Larochelle, Laviolette, Marchand, and Lempitsky]{ganin2016domain}
Yaroslav Ganin, Evgeniya Ustinova, Himan Ajakan, Pascal Germain, Hugo Larochelle, Fran{\c{c}}ois Laviolette, Mario Marchand, and Victor Lempitsky.
\newblock Domain-adversarial training of neural networks.
\newblock In \emph{Proceedings of the 33rd International Conference on Machine Learning (ICML)}, pp.\  2972--2981, 2016.

\bibitem[Gehman et~al.(2020)Gehman, Gururangan, Sap, and Choi]{gehman2020realtotoxicityprompts}
Samuel Gehman, Suchin Gururangan, Maarten Sap, and Yejin Choi.
\newblock Realtoxicityprompts: Evaluating neural toxic degeneration in language models.
\newblock In \emph{Proceedings of the 2020 Conference on Empirical Methods in Natural Language Processing (EMNLP)}, pp.\  3403--3413, 2020.

\bibitem[Goodfellow et~al.(2015)Goodfellow, Shlens, and Szegedy]{goodfellow2014explaining}
Ian~J. Goodfellow, Jonathon Shlens, and Christian Szegedy.
\newblock Explaining and harnessing adversarial examples.
\newblock In \emph{International Conference on Learning Representations (ICLR)}, 2015.

\bibitem[Hausenloy et~al.(2024)Hausenloy, McClements, and Thakur]{hausenloy2024datagovernancefrontierai}
Jason Hausenloy, Duncan McClements, and Madhavendra Thakur.
\newblock Towards data governance of frontier ai models, 2024.
\newblock URL \url{https://arxiv.org/abs/2412.03824}.

\bibitem[Jagielski et~al.(2018)Jagielski, Oprea, Biggio, et~al.]{jagielski2018manipulating}
Marcin Jagielski, Alexandru Oprea, Battista Biggio, et~al.
\newblock Manipulating machine learning: Poisoning attacks and countermeasures for regression.
\newblock In \emph{Proceedings of the International Conference on Machine Learning (ICML)}, pp.\  2819--2828, 2018.

\bibitem[Karras et~al.(2019)Karras, Laine, and Aila]{karras2019style}
Tero Karras, Samuli Laine, and Timo Aila.
\newblock A style-based generator architecture for generative adversarial networks.
\newblock In \emph{Proceedings of the IEEE Conference on Computer Vision and Pattern Recognition (CVPR)}, pp.\  4401--4410, 2019.

\bibitem[Kingma \& Welling(2014)Kingma and Welling]{kingma2013auto}
Diederik~P. Kingma and Max Welling.
\newblock Auto-encoding variational bayes.
\newblock In \emph{International Conference on Learning Representations (ICLR)}, 2014.

\bibitem[Kumar et~al.(2022)Kumar, Hines, and Dickerson]{Kumar_2022}
I.~Elizabeth Kumar, Keegan~E. Hines, and John~P. Dickerson.
\newblock Equalizing credit opportunity in algorithms: Aligning algorithmic fairness research with u.s. fair lending regulation.
\newblock In \emph{Proceedings of the 2022 AAAI/ACM Conference on AI, Ethics, and Society}, AIES ’22, pp.\  357–368. ACM, July 2022.
\newblock \doi{10.1145/3514094.3534154}.
\newblock URL \url{http://dx.doi.org/10.1145/3514094.3534154}.

\bibitem[Madry et~al.(2018)Madry, Makelov, Schmidt, Tsipras, and Vladu]{madry2018towards}
Aleksander Madry, Aleksandar Makelov, Ludwig Schmidt, Dimitris Tsipras, and Adrian Vladu.
\newblock Towards deep learning models resistant to adversarial attacks.
\newblock In \emph{International Conference on Learning Representations (ICLR)}, 2018.

\bibitem[Mehrabi et~al.(2021)Mehrabi, Morstatter, Saxena, Lerman, and Galstyan]{mehrabi2021survey}
Ninareh Mehrabi, Fred Morstatter, Nripsuta Saxena, Kale Lerman, and Aram Galstyan.
\newblock A survey on bias and fairness in machine learning.
\newblock \emph{ACM Computing Surveys (CSUR)}, 54\penalty0 (6):\penalty0 1--35, 2021.

\bibitem[OpenAI et~al.(2024)OpenAI, Achiam, Adler, Agarwal, Ahmad, Akkaya, Aleman, Almeida, Altenschmidt, Altman, Anadkat, Avila, Babuschkin, Balaji, Balcom, Baltescu, Bao, Bavarian, Belgum, Bello, Berdine, Bernadett-Shapiro, Berner, Bogdonoff, Boiko, Boyd, Brakman, Brockman, Brooks, Brundage, Button, Cai, Campbell, Cann, Carey, Carlson, Carmichael, Chan, Chang, Chantzis, Chen, Chen, Chen, Chen, Chen, Chess, Cho, Chu, Chung, Cummings, Currier, Dai, Decareaux, Degry, Deutsch, Deville, Dhar, Dohan, Dowling, Dunning, Ecoffet, Eleti, Eloundou, Farhi, Fedus, Felix, Fishman, Forte, Fulford, Gao, Georges, Gibson, Goel, Gogineni, Goh, Gontijo-Lopes, Gordon, Grafstein, Gray, Greene, Gross, Gu, Guo, Hallacy, Han, Harris, He, Heaton, Heidecke, Hesse, Hickey, Hickey, Hoeschele, Houghton, Hsu, Hu, Hu, Huizinga, Jain, Jain, Jang, Jiang, Jiang, Jin, Jin, Jomoto, Jonn, Jun, Kaftan, Łukasz Kaiser, Kamali, Kanitscheider, Keskar, Khan, Kilpatrick, Kim, Kim, Kim, Kirchner, Kiros, Knight, Kokotajlo, Łukasz Kondraciuk, Kondrich,
  Konstantinidis, Kosic, Krueger, Kuo, Lampe, Lan, Lee, Leike, Leung, Levy, Li, Lim, Lin, Lin, Litwin, Lopez, Lowe, Lue, Makanju, Malfacini, Manning, Markov, Markovski, Martin, Mayer, Mayne, McGrew, McKinney, McLeavey, McMillan, McNeil, Medina, Mehta, Menick, Metz, Mishchenko, Mishkin, Monaco, Morikawa, Mossing, Mu, Murati, Murk, Mély, Nair, Nakano, Nayak, Neelakantan, Ngo, Noh, Ouyang, O'Keefe, Pachocki, Paino, Palermo, Pantuliano, Parascandolo, Parish, Parparita, Passos, Pavlov, Peng, Perelman, de~Avila Belbute~Peres, Petrov, de~Oliveira~Pinto, Michael, Pokorny, Pokrass, Pong, Powell, Power, Power, Proehl, Puri, Radford, Rae, Ramesh, Raymond, Real, Rimbach, Ross, Rotsted, Roussez, Ryder, Saltarelli, Sanders, Santurkar, Sastry, Schmidt, Schnurr, Schulman, Selsam, Sheppard, Sherbakov, Shieh, Shoker, Shyam, Sidor, Sigler, Simens, Sitkin, Slama, Sohl, Sokolowsky, Song, Staudacher, Such, Summers, Sutskever, Tang, Tezak, Thompson, Tillet, Tootoonchian, Tseng, Tuggle, Turley, Tworek, Uribe, Vallone, Vijayvergiya,
  Voss, Wainwright, Wang, Wang, Wang, Ward, Wei, Weinmann, Welihinda, Welinder, Weng, Weng, Wiethoff, Willner, Winter, Wolrich, Wong, Workman, Wu, Wu, Wu, Xiao, Xu, Yoo, Yu, Yuan, Zaremba, Zellers, Zhang, Zhang, Zhao, Zheng, Zhuang, Zhuk, and Zoph]{openai2024gpt4technicalreport}
OpenAI, Josh Achiam, Steven Adler, Sandhini Agarwal, Lama Ahmad, Ilge Akkaya, Florencia~Leoni Aleman, Diogo Almeida, Janko Altenschmidt, Sam Altman, Shyamal Anadkat, Red Avila, Igor Babuschkin, Suchir Balaji, Valerie Balcom, Paul Baltescu, Haiming Bao, Mohammad Bavarian, Jeff Belgum, Irwan Bello, Jake Berdine, Gabriel Bernadett-Shapiro, Christopher Berner, Lenny Bogdonoff, Oleg Boiko, Madelaine Boyd, Anna-Luisa Brakman, Greg Brockman, Tim Brooks, Miles Brundage, Kevin Button, Trevor Cai, Rosie Campbell, Andrew Cann, Brittany Carey, Chelsea Carlson, Rory Carmichael, Brooke Chan, Che Chang, Fotis Chantzis, Derek Chen, Sully Chen, Ruby Chen, Jason Chen, Mark Chen, Ben Chess, Chester Cho, Casey Chu, Hyung~Won Chung, Dave Cummings, Jeremiah Currier, Yunxing Dai, Cory Decareaux, Thomas Degry, Noah Deutsch, Damien Deville, Arka Dhar, David Dohan, Steve Dowling, Sheila Dunning, Adrien Ecoffet, Atty Eleti, Tyna Eloundou, David Farhi, Liam Fedus, Niko Felix, Simón~Posada Fishman, Juston Forte, Isabella Fulford, Leo
  Gao, Elie Georges, Christian Gibson, Vik Goel, Tarun Gogineni, Gabriel Goh, Rapha Gontijo-Lopes, Jonathan Gordon, Morgan Grafstein, Scott Gray, Ryan Greene, Joshua Gross, Shixiang~Shane Gu, Yufei Guo, Chris Hallacy, Jesse Han, Jeff Harris, Yuchen He, Mike Heaton, Johannes Heidecke, Chris Hesse, Alan Hickey, Wade Hickey, Peter Hoeschele, Brandon Houghton, Kenny Hsu, Shengli Hu, Xin Hu, Joost Huizinga, Shantanu Jain, Shawn Jain, Joanne Jang, Angela Jiang, Roger Jiang, Haozhun Jin, Denny Jin, Shino Jomoto, Billie Jonn, Heewoo Jun, Tomer Kaftan, Łukasz Kaiser, Ali Kamali, Ingmar Kanitscheider, Nitish~Shirish Keskar, Tabarak Khan, Logan Kilpatrick, Jong~Wook Kim, Christina Kim, Yongjik Kim, Jan~Hendrik Kirchner, Jamie Kiros, Matt Knight, Daniel Kokotajlo, Łukasz Kondraciuk, Andrew Kondrich, Aris Konstantinidis, Kyle Kosic, Gretchen Krueger, Vishal Kuo, Michael Lampe, Ikai Lan, Teddy Lee, Jan Leike, Jade Leung, Daniel Levy, Chak~Ming Li, Rachel Lim, Molly Lin, Stephanie Lin, Mateusz Litwin, Theresa Lopez, Ryan
  Lowe, Patricia Lue, Anna Makanju, Kim Malfacini, Sam Manning, Todor Markov, Yaniv Markovski, Bianca Martin, Katie Mayer, Andrew Mayne, Bob McGrew, Scott~Mayer McKinney, Christine McLeavey, Paul McMillan, Jake McNeil, David Medina, Aalok Mehta, Jacob Menick, Luke Metz, Andrey Mishchenko, Pamela Mishkin, Vinnie Monaco, Evan Morikawa, Daniel Mossing, Tong Mu, Mira Murati, Oleg Murk, David Mély, Ashvin Nair, Reiichiro Nakano, Rajeev Nayak, Arvind Neelakantan, Richard Ngo, Hyeonwoo Noh, Long Ouyang, Cullen O'Keefe, Jakub Pachocki, Alex Paino, Joe Palermo, Ashley Pantuliano, Giambattista Parascandolo, Joel Parish, Emy Parparita, Alex Passos, Mikhail Pavlov, Andrew Peng, Adam Perelman, Filipe de~Avila Belbute~Peres, Michael Petrov, Henrique~Ponde de~Oliveira~Pinto, Michael, Pokorny, Michelle Pokrass, Vitchyr~H. Pong, Tolly Powell, Alethea Power, Boris Power, Elizabeth Proehl, Raul Puri, Alec Radford, Jack Rae, Aditya Ramesh, Cameron Raymond, Francis Real, Kendra Rimbach, Carl Ross, Bob Rotsted, Henri Roussez,
  Nick Ryder, Mario Saltarelli, Ted Sanders, Shibani Santurkar, Girish Sastry, Heather Schmidt, David Schnurr, John Schulman, Daniel Selsam, Kyla Sheppard, Toki Sherbakov, Jessica Shieh, Sarah Shoker, Pranav Shyam, Szymon Sidor, Eric Sigler, Maddie Simens, Jordan Sitkin, Katarina Slama, Ian Sohl, Benjamin Sokolowsky, Yang Song, Natalie Staudacher, Felipe~Petroski Such, Natalie Summers, Ilya Sutskever, Jie Tang, Nikolas Tezak, Madeleine~B. Thompson, Phil Tillet, Amin Tootoonchian, Elizabeth Tseng, Preston Tuggle, Nick Turley, Jerry Tworek, Juan Felipe~Cerón Uribe, Andrea Vallone, Arun Vijayvergiya, Chelsea Voss, Carroll Wainwright, Justin~Jay Wang, Alvin Wang, Ben Wang, Jonathan Ward, Jason Wei, CJ~Weinmann, Akila Welihinda, Peter Welinder, Jiayi Weng, Lilian Weng, Matt Wiethoff, Dave Willner, Clemens Winter, Samuel Wolrich, Hannah Wong, Lauren Workman, Sherwin Wu, Jeff Wu, Michael Wu, Kai Xiao, Tao Xu, Sarah Yoo, Kevin Yu, Qiming Yuan, Wojciech Zaremba, Rowan Zellers, Chong Zhang, Marvin Zhang, Shengjia
  Zhao, Tianhao Zheng, Juntang Zhuang, William Zhuk, and Barret Zoph.
\newblock Gpt-4 technical report, 2024.
\newblock URL \url{https://arxiv.org/abs/2303.08774}.

\bibitem[Reuel et~al.(2024)Reuel, Bucknall, Casper, Fist, Soder, Aarne, Hammond, Ibrahim, Chan, Wills, Anderljung, Garfinkel, Heim, Trask, Mukobi, Schaeffer, Baker, Hooker, Solaiman, Luccioni, Rajkumar, Moës, Ladish, Guha, Newman, Bengio, South, Pentland, Koyejo, Kochenderfer, and Trager]{reuel2024openproblemstechnicalai}
Anka Reuel, Ben Bucknall, Stephen Casper, Tim Fist, Lisa Soder, Onni Aarne, Lewis Hammond, Lujain Ibrahim, Alan Chan, Peter Wills, Markus Anderljung, Ben Garfinkel, Lennart Heim, Andrew Trask, Gabriel Mukobi, Rylan Schaeffer, Mauricio Baker, Sara Hooker, Irene Solaiman, Alexandra~Sasha Luccioni, Nitarshan Rajkumar, Nicolas Moës, Jeffrey Ladish, Neel Guha, Jessica Newman, Yoshua Bengio, Tobin South, Alex Pentland, Sanmi Koyejo, Mykel~J. Kochenderfer, and Robert Trager.
\newblock Open problems in technical ai governance, 2024.
\newblock URL \url{https://arxiv.org/abs/2407.14981}.

\bibitem[Richter et~al.(2020)Richter, Roy, and Heger]{richter2020can}
Sven Richter, Daniel Roy, and Daniel Heger.
\newblock Can synthetic data bridge the reality gap in autonomous driving?
\newblock In \emph{Proceedings of the IEEE/CVF Conference on Computer Vision and Pattern Recognition Workshops (CVPRW)}, pp.\  100--105, 2020.

\bibitem[Ruan et~al.(2024)Ruan, Maddison, and Hashimoto]{ruan2024observationalscalinglawspredictability}
Yangjun Ruan, Chris~J. Maddison, and Tatsunori Hashimoto.
\newblock Observational scaling laws and the predictability of language model performance, 2024.
\newblock URL \url{https://arxiv.org/abs/2405.10938}.

\bibitem[Sastry et~al.(2024)Sastry, Heim, Belfield, Anderljung, Brundage, Hazell, O'Keefe, Hadfield, Ngo, Pilz, Gor, Bluemke, Shoker, Egan, Trager, Avin, Weller, Bengio, and Coyle]{sastry2024computingpowergovernanceartificial}
Girish Sastry, Lennart Heim, Haydn Belfield, Markus Anderljung, Miles Brundage, Julian Hazell, Cullen O'Keefe, Gillian~K. Hadfield, Richard Ngo, Konstantin Pilz, George Gor, Emma Bluemke, Sarah Shoker, Janet Egan, Robert~F. Trager, Shahar Avin, Adrian Weller, Yoshua Bengio, and Diane Coyle.
\newblock Computing power and the governance of artificial intelligence, 2024.
\newblock URL \url{https://arxiv.org/abs/2402.08797}.

\bibitem[Shrivastava et~al.(2017)Shrivastava, Pfister, Tuzel, Susskind, Wang, and Webb]{shrivastava2017learning}
Anurag Shrivastava, Tomas Pfister, {\"O}ncel Tuzel, Joshua Susskind, Wenzhe Wang, and Rob Webb.
\newblock Learning from simulated and unsupervised images through adversarial training.
\newblock In \emph{Proceedings of the IEEE Conference on Computer Vision and Pattern Recognition (CVPR)}, pp.\  2242--2251, 2017.

\bibitem[Steinhardt et~al.(2017)Steinhardt, Koh, and Liang]{steinhardt2017certified}
Jacob Steinhardt, Pang~Wei Koh, and Percy Liang.
\newblock Certified defenses for data poisoning attacks.
\newblock In \emph{Advances in Neural Information Processing Systems}, volume~30, pp.\  3517--3529, 2017.

\bibitem[Torralba \& Efros(2011)Torralba and Efros]{torralba2011unbiased}
Antonio Torralba and Alexei~C. Efros.
\newblock Unbiased look at dataset bias.
\newblock In \emph{Proceedings of the IEEE Conference on Computer Vision and Pattern Recognition (CVPR)}, pp.\  1521--1528, 2011.

\bibitem[Villalobos et~al.(2024)Villalobos, Ho, Sevilla, Besiroglu, Heim, and Hobbhahn]{villalobos2024rundatalimitsllm}
Pablo Villalobos, Anson Ho, Jaime Sevilla, Tamay Besiroglu, Lennart Heim, and Marius Hobbhahn.
\newblock Will we run out of data? limits of llm scaling based on human-generated data, 2024.
\newblock URL \url{https://arxiv.org/abs/2211.04325}.

\bibitem[Wei \& Zou(2019)Wei and Zou]{wei2019eda}
Jason Wei and Kai Zou.
\newblock Eda: Easy data augmentation techniques for boosting performance on text classification tasks.
\newblock In \emph{Proceedings of the 2019 Conference on Empirical Methods in Natural Language Processing and the 9th International Joint Conference on Natural Language Processing (EMNLP-IJCNLP)}, pp.\  6381--6387, 2019.

\bibitem[Xu et~al.(2019)Xu, Zhang, Liu, Zhou, and Zheng]{xu2019fairgan}
Yang Xu, Yifu Zhang, Ming Liu, Haotian Zhou, and Wei Zheng.
\newblock Fairgan: Fairness-aware generative adversarial networks.
\newblock In \emph{Proceedings of the AAAI Conference on Artificial Intelligence}, volume~33, pp.\  10267--10274, 2019.

\bibitem[Ziegler et~al.(2019)Ziegler, Stiennon, Wu, Brown, Radford, Amodei, and Christiano]{ziegler2019fine}
Daniel~M Ziegler, Nicolas Stiennon, Jeffrey Wu, Tom~B Brown, Alec Radford, Dario Amodei, and Paul~F Christiano.
\newblock Fine-tuning language models from human preferences.
\newblock \emph{arXiv preprint arXiv:1909.08593}, 2019.

\end{thebibliography}
\bibliographystyle{iclr2025_conference}

\end{document}